\documentclass[11pt]{article}

\usepackage[T1]{fontenc}
\usepackage{lmodern}
\usepackage[margin=1in]{geometry}
\usepackage{graphicx}
\usepackage{amsmath,amssymb}
\usepackage[hidelinks]{hyperref}

\title{GeoDiff-SAR: A Geometric-Prior-Guided Diffusion Model for SAR Image Generation}

\author{
  \footnotesize Fan Zhang$^{1}$, Xuanting Wu$^{1}$, and Fei Ma$^{1}$\thanks{Corresponding author: \href{mailto:mafei@mail.buct.edu.cn}{mafei@mail.buct.edu.cn}.}\\
  \footnotesize Qiang Yin$^{1}$ and Yuxin Hu$^{2}$\\[0.5em]
  \footnotesize $^{1}$College of Information Science and Technology,\\
  \footnotesize Beijing University of Chemical Technology, Beijing 100029, China\\
  \footnotesize $^{2}$Aerospace Information Research Institute, Chinese Academy of Sciences, Beijing 100190, China
}
\date{}

\begin{document}
\maketitle

\begin{abstract}
Synthetic aperture radar (SAR) image generation can mitigate data scarcity, but controllable generation under sparse observation angles remains difficult. Recent SAR generative studies improve texture realism, yet explicit geometry-aware control is still limited. This paper studies the focused and verifiable setting of intermediate-azimuth completion: 3D-model-derived geometric priors guide a diffusion model to synthesize the views missing from sparse-angle training data. GeoDiff-SAR constructs a lightweight multi-bounce ray-tracing prior, encodes the resulting point cloud, and fuses it with text conditioning while adapting Stable Diffusion 3.5 Medium through low-rank adaptation. On a real four-category aircraft dataset, GeoDiff-SAR reaches an SSIM of 0.812 and azimuth consistency of 0.940, compared with 0.738 and 0.782 for the text-conditioned SD3.5 Medium baseline. The same sparse-angle protocol on five MSTAR vehicle classes yields an SSIM of 0.878 and azimuth consistency of 0.917. These results support the conclusion that a lightweight 3D geometric prior improves viewpoint adherence for controllable SAR generation; it is intended as generation guidance rather than high-fidelity electromagnetic reconstruction.
\end{abstract}

\section{Introduction}

Synthetic aperture radar (SAR) image generation is increasingly used to alleviate the limited availability of labeled SAR data. Although recent generative methods improve texture realism, explicit geometry-aware control remains limited \cite{huang2024survey,zhang2025phgan}. The difficulty is particularly pronounced for aircraft SAR: changes in azimuth reshape layover, shadow, and the dominant scattering centers. We therefore focus on a narrower but testable claim than arbitrary continuous-angle synthesis: a 3D-model-derived prior can enable controllable intermediate-azimuth completion when the model is trained only at sparse angles.

\begin{figure*}[t]
    \centering
    \includegraphics[width=\textwidth]{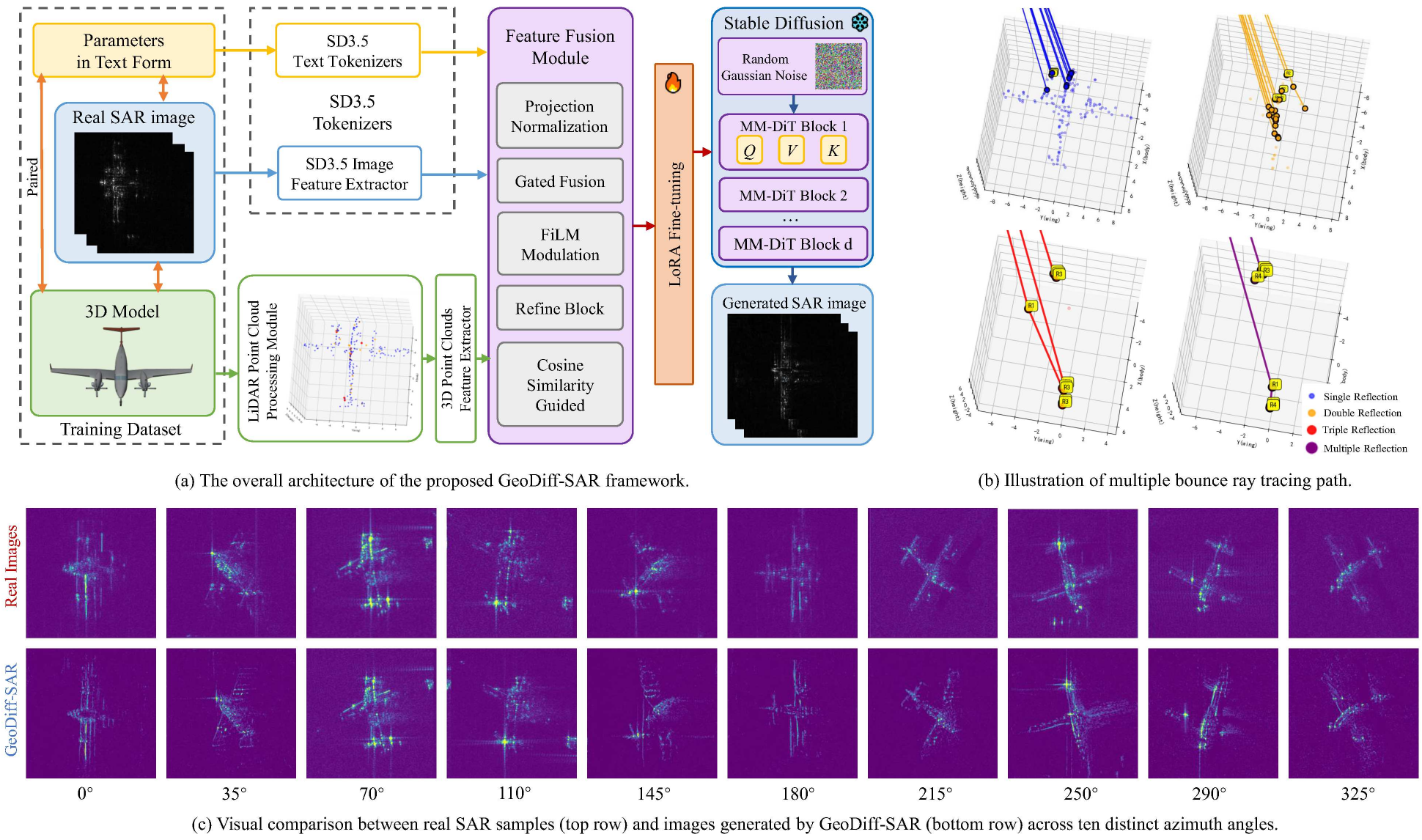}
    \caption{Overview and visual evidence of GeoDiff-SAR. (a) The 3D-model-derived point-cloud prior is fused with textual conditions to guide diffusion-based SAR generation. (b) Multi-bounce ray-tracing paths construct the geometric prior. (c) Real SAR samples (top) and GeoDiff-SAR results (bottom) across representative azimuths. The generated images preserve view-dependent structures and dominant scattering patterns under sparse-angle training.}
    \label{fig:overall}
\end{figure*}

Figure~\ref{fig:overall} gives an overview of GeoDiff-SAR and the evidence considered here. A text prompt specifies the target, radar setting, and requested azimuth; for example, ``SAR image, Pilatus PC 12, Ku-band, 0.5 m resolution, VV polarization, azimuth 5 degree.'' The system retrieves the associated 3D model and constructs a viewpoint-specific geometric prior. Measured azimuths that do not fall on the $5^\circ$ grid are quantized to the nearest discrete azimuth label for prompting. The ray-casting branch is deliberately lightweight: it supplies geometry-aware scattering cues rather than a rigorous electromagnetic simulation \cite{auer2010ray}. The prior is encoded and fused with the native text-conditioning branches of Stable Diffusion 3.5 Medium, which are adapted to the SAR domain using low-rank adaptation (LoRA) \cite{esser2024sd3,hu2022lora}.

\section{Geometry-Guided Diffusion}

\subsection{Geometric Prior Construction}

The geometric branch simulates multi-bounce scattering cues and occlusion relationships by recursive ray tracing. Its fixed heuristic parameters are $\alpha_{\mathrm{edge}}=2.0$, $\alpha_{\mathrm{vert}}=2.0$, $\alpha_{\mathrm{struct}}=1.6$, $\tau_{\mathrm{area}}=0.01$, $\tau_{\mathrm{vert}}=0.3$, $\zeta=0.8$, $\mu=0.03$, $K_{\max}=3$, $\tau_{\min}=0.03$, $\sigma_{\mathrm{scan}}=0.3$, and $\sigma_{\mathrm{reflect}}=0.1$. These values are manually selected once for the acquisition setting and then fixed across categories, polarizations, and viewpoints.

For ray state $S_k=(\mathbf{p}_k,\mathbf{d}_k,E_k)$, the scattering intensity is approximated by
\begin{equation}
I_{\mathrm{scatter}}=I_k e^{-\mu L_{\mathrm{path}}}
\mathcal{W}_{\mathrm{base}}\mathcal{H}_{\mathrm{edge}}
\mathcal{H}_{\mathrm{orient}}\mathcal{H}_{\mathrm{struct}},
\end{equation}
where the factors respectively encode the base response, edge enhancement, side-looking orientation, and structural scattering. The reflected direction is updated using
\begin{equation}
\mathbf{d}_{k+1}=\operatorname{Normalize}(\mathbf{d}_{\mathrm{spec}}+\zeta\mathbf{u}),
\qquad \mathbf{u}\sim\mathcal{U}(-1,1)^3,
\end{equation}
which adds a roughness-controlled diffuse component. The result is a geometry-aware point cloud with approximate scattering and visibility cues. It is a practical condition generator for sparse-angle synthesis, not a substitute for full electromagnetic reconstruction.

\subsection{Condition Fusion and Generation}

The point-cloud prior is fused with text features and, during training, an auxiliary image feature by adaptive weighted fusion,
\begin{equation}
C_{\mathrm{fused}}=\alpha_t\tilde{F}_t+\alpha_p\tilde{F}_p+\alpha_i\tilde{F}_i,
\qquad \alpha_t+\alpha_p+\alpha_i=1,
\end{equation}
where $\tilde{F}_t$, $\tilde{F}_p$, and $\tilde{F}_i$ denote text, geometric, and auxiliary image features. The image feature is a training-time visual anchor that stabilizes texture--semantic alignment; it is unavailable and omitted during inference. Generation is therefore driven by text plus geometry, without requiring a paired SAR image.

We adapt Stable Diffusion 3.5 Medium with LoRA while retaining most of the backbone. The diffusion model uses the usual noise-prediction loss,
\begin{equation}
\mathcal{L}_{\mathrm{simple}}=\mathbb{E}\left[\left\|\epsilon-\epsilon_{\theta}(z_t,t,C_{\mathrm{fused}})\right\|_2^2\right].
\end{equation}

\section{Experimental Protocol}

\subsection{Aircraft Dataset and Sparse-Angle Evaluation}

The real SAR aircraft dataset contains 8,536 images from four categories: Cessna 208 (3,468), Kodiak 100 (3,468), King Air 350i (800), and Pilatus PC 12 (800). It covers four polarizations and azimuths from $0^\circ$ to $355^\circ$ at $5^\circ$ intervals. The training split contains only 1,870 samples at $10^\circ$ intervals, while the 6,666-image test split retains full $5^\circ$ sampling.

We evaluate intermediate-azimuth completion: the model trains with $0^\circ,10^\circ,20^\circ,\ldots$ observations and generates the missing $5^\circ,15^\circ,25^\circ,\ldots$ views. Quantitative comparisons use real held-out SAR images with the same semantic conditions. As a public-benchmark sanity check, we also evaluate five MSTAR vehicle classes (2S1, BMP2, BTR70, D7, and T72) at fixed $17^\circ$ depression and HH polarization. This split contains 677 training samples at $10^\circ$ intervals and 1,581 test samples at $5^\circ$ intervals. The geometric-prior parameters are unchanged.

\subsection{Metrics and Results}

Natural-image perceptual metrics are imperfect proxies for SAR quality, so we emphasize structural similarity (SSIM) and azimuth consistency. Azimuth consistency is
\begin{equation}
\mathrm{Acc}_{\mathrm{azi}}=\frac{1}{N}\sum_{n=1}^{N}\mathbf{1}\!\left[\hat{y}_n=y_n\right],
\end{equation}
where $y_n$ is the commanded azimuth label and $\hat{y}_n$ is predicted by an azimuth evaluator trained only on real SAR images.

On the aircraft set, GeoDiff-SAR achieves SSIM $=0.812$ and azimuth consistency $=0.940$, compared with $0.738$ and $0.782$ for SD3.5 Medium without the 3D geometric prior. On MSTAR, the framework reaches SSIM $=0.878$ and azimuth consistency $=0.917$, exceeding the baseline values of $0.733$ and $0.804$. Together with the visual comparisons in Figure~\ref{fig:overall}, these improvements indicate better viewpoint adherence rather than only SAR-like texture synthesis.

\subsection{Sensitivity and Efficiency}

With $\zeta=0.8$, varying $K_{\max}\in\{1,2,3\}$ yields (SSIM, azimuth consistency) of $(0.736,0.824)$, $(0.784,0.866)$, and $(0.812,0.940)$. With $K_{\max}=3$, varying $\zeta\in\{0,0.4,0.8\}$ yields $(0.739,0.871)$, $(0.758,0.902)$, and $(0.812,0.940)$. We retain $K_{\max}=3$ and $\zeta=0.8$. Averaged over 100 trials, prior construction takes approximately 48 seconds per azimuth on CPU; diffusion inference takes approximately 13 seconds per image on an RTX 4090 with 30 sampling steps. Thus, geometric-prior construction is the main additional computational cost.

\section{Discussion and Conclusion}

The evidence supports a focused conclusion: 3D-model-derived geometric priors can guide SAR generation under sparse-angle training and improve controllable intermediate-azimuth completion. We do not claim arbitrary continuous-angle synthesis or strict physical reconstruction. The prior is heuristic and should be understood as a lightweight guidance mechanism rather than a high-fidelity electromagnetic solver. Although the main high-resolution aircraft dataset is not public, the MSTAR check shows that the approach retains gains on a public benchmark. Broader cross-sensor and cross-scene validation remains future work.

\section*{Acknowledgments}

This work was supported by the National Natural Science Foundation of China (Grant Nos. 62201027 and 62271034).

\end{document}